\documentclass[letter,scriptaddress,noshowkeys,showpacs,superscriptaddress,notitlepage]{revtex4-1}

	\usepackage{natbib}
	\usepackage{amsmath}%,amssymb} 
	\usepackage{makeidx}
	\usepackage{amsfonts}
	\usepackage[ansinew]{inputenc}
	\usepackage[usenames,dvipsnames]{pstricks}
	\usepackage{subfigure}
	\usepackage{epsfig}
	\usepackage{pst-grad} % For gradients
	\usepackage{pst-plot} % For axes
	\usepackage[colorlinks,hyperindex]{hyperref}
	\usepackage{epstopdf}
	\usepackage{multirow}
	\usepackage{titlesec}
	\usepackage{float}

	\titlespacing{\section}{0pt}{5pt}{5pt}
	\titlespacing{\subsection}{0pt}{5pt}{5pt}
	\titlespacing{\subsubsection}{0pt}{5pt}{5pt}

	\newcommand{\squeezeup}{\vspace{-5mm}}

	\hypersetup
	{
		colorlinks,%
		citecolor=black,%
		linkcolor=black,%
		urlcolor=black,%
	}

\makeindex

\begin{document}

\title{Near-space flight of a correlated photon system}

\author{Zhongkan Tang}
\affiliation{Centre for Quantum Technologies, National University of Singapore,\\ Block S15, 3 Science Drive 2, 117543 Singapore.}
\author{Rakhitha Chandrasekara}
\affiliation{Centre for Quantum Technologies, National University of Singapore,\\ Block S15, 3 Science Drive 2, 117543 Singapore.}
\author{Yau Yong Sean}
\affiliation{Centre for Quantum Technologies, National University of Singapore,\\ Block S15, 3 Science Drive 2, 117543 Singapore.}
\author{Cliff Cheng}
\affiliation{Centre for Quantum Technologies, National University of Singapore,\\ Block S15, 3 Science Drive 2, 117543 Singapore.}
\author{Christoph Wildfeuer}
\affiliation{Kantonsschule Sursee, \\ Moosgasse 11, CH-6210 Sursee, Switzerland.}
\author{Alexander Ling}
\affiliation{Centre for Quantum Technologies, National University of Singapore,\\ Block S15, 3 Science Drive 2, 117543 Singapore.}

\date{15 April 2014}

\begin{abstract}
We report the successful test flight of a device for generating and monitoring correlated photon pairs under near-space conditions up to 35.5\,km altitude. 
Data from ground based qualification tests and the high altitude experiment demonstrate that the device continues to operate even under harsh environmental conditions. 
The design of the rugged, compact and power-efficient photon pair system is presented. 
This design enables autonomous photon pair systems to be deployed on low-resource platforms such as nanosatellites hosting remote nodes of a quantum key distribution network.
These results pave the way for tests of entangled photon technology in low earth orbit.
\end{abstract}

\keywords{nanosatellites, correlated photon sources, space qualification, quantum key distribution}

\maketitle

\section{Introduction}

Correlated photon pair sources based on spontaneous parametric down conversion (SPDC) are an established technology \cite{coinc70}.
Since pioneering demonstrations the brightness and quality of the photon pair correlations have been steadily improving \cite{ou88, kwiat95, kwiat99, kurtsiefer01, trojek08, steinlechner12}.
The most popular application of SPDC sources is the generation of photon pairs described by a quantum entangled state. 

Ongoing research on SPDC sources is largely spurred by interest in quantum communication. In particular, quantum key distribution (QKD) based on entangled photon pairs continues to attract attention due to superior privacy guarantees provided by fundamental quantum mechanics \cite{E91,gisin02,qcomm07,acin07}.  
Entanglement distribution can take place via optical fiber or free-space links but current fiber technology imposes a practical distance limit due to loss and decoherence.  
These limitations suggest that QKD networks beyond metropolitan-scale distances will rely on free-space links \cite{ursin07}. 
By coupling photon pair sources to optical transmitters and placing them aboard high altitude air- or space-craft, it is possible to beam entangled photons to widely separated receivers.
Research on planet-wide entanglement distribution using space-craft is ongoing \cite{PWQuantumSpace} and a major milestone in this direction would be the demonstration of a bright entangled photon pair source in low earth orbit (LEO).

We propose that an entangled photon source can be demonstrated cost effectively in LEO using small space-craft called nanosatellites \cite{morong12}.
In particular the CubeSat standard \cite{woellert11} is very attractive due to its short design cycle, the use of commercial-off-the-shelf (COTS) components and a common deployment mechanism. The standard CubeSat unit is a 1\,kg, 100\,mm cube (1U), and these units can be stacked to build larger space-craft. 

The disadvantage of using CubeSats is the limited availability of resources.
In a typical 1U CubeSat limits on the payload are the form factor (100$\times$100$\times$30\,mm$^3$), mass (300\,gm) and power (2\,W continuous).
Other challenges include mechanical vibration experienced during a rocket launch, lack of shielding against radiation, possible thermal fatigue due to lack of insulation material and limited power for temperature regulation.

In response to these challenges we have designed a small, rugged and power efficient system for {\it generating and monitoring} polarization correlations between photon pairs. 
The system is packaged into a device whose form factor and power requirements are consistent with the payload limits of a 1U CubeSat.
Within the package are closely integrated optical and electronic sub-systems.
The electronics segment has the complete infrastructure for carrying out photon pair counting experiments and is fully automated.

The source emits classically correlated photon pairs based on type I collinear SPDC in a single crystal.
To validate the key parameters of ruggedness and power efficiency it is sufficient to generate and monitor strong {\it classical} correlation between pairs of photons. 
If the design is able to support a basic SPDC source it can be extended to generate polarization entangled photon pairs \cite{trojek08}.

%We present the design of the optical and electronic sub-systems followed by a description of the qualification tests that the device was subjected to.  
%We conclude with the results from our test in near-space conditions. 

\section{The optical sub-system}

\begin{figure}[htbp]
	\includegraphics[scale=0.9]{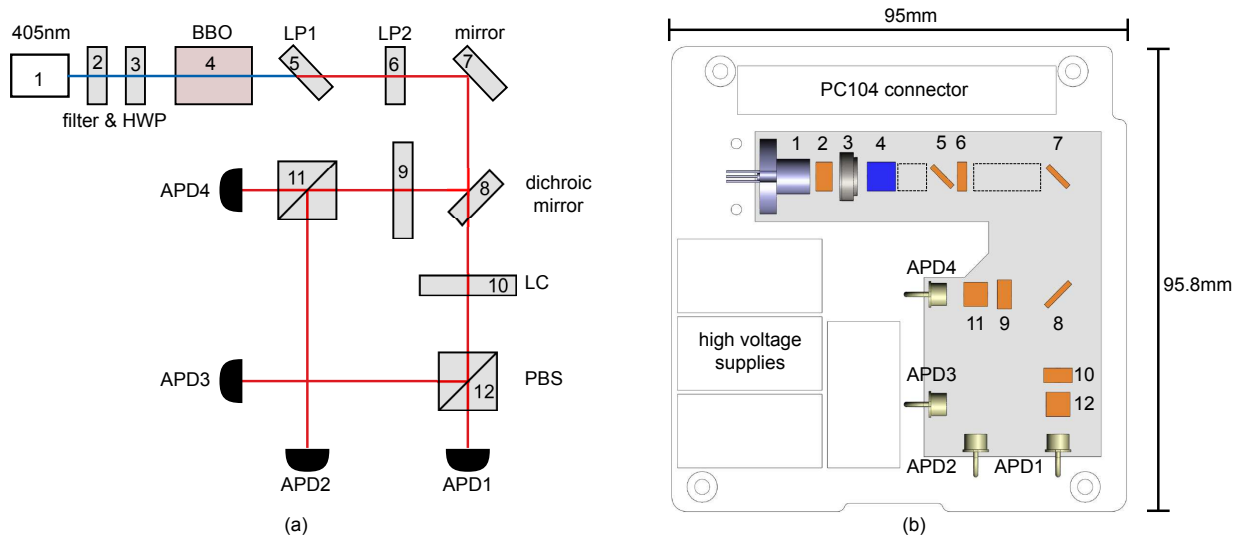}
	\caption{Schematic of elements in the optical subsystem. Fig.1(a) Correlated photon pairs at 760 and 867\,nm are generated through {\it $\beta$}-barium-borate (BBO) via type I SPDC pumped by a 405\,nm laser diode. 
The pump spectrum and polarization mode is prepared by an appropriate band-pass filter and half-wave plate (HWP).
Excess pump light is removed via long pass filters (LP1 \& LP2).
The liquid crystal (LC) units and polarizing beam splitters (PBS) are used for rotating and analyzing photon pairs. Two pairs of silicon avalanche photodiodes (APDs) provide redundancy in the detectors. Each APD has individual spatial and spectral filters and an individual high voltage supply. The optical unit is placed on a printed circuit board. Fig.1(b) is drawn to scale with 1:1.7 ratio, showing the position of the essential elements within the aluminium housing (colored grey online). These are labeled by number for cross referencing. The dashed boxes are placeholders for additional crystals to enable production of polarization entangled photon pairs \cite{trojek08}. The electronic sub-system is placed on the reverse side of the board. The completed device does not exceed 30\,mm in height.}
	\label{opticssetup}
	\squeezeup
	\end{figure}
The photon pair source and detectors are housed within a machined aluminium optical unit that is black anodized to minimize internal reflection. The layout is shown to scale in Fig.\,\ref{opticssetup}. 
The optical pump is a continuous wave single mode 405\,nm laser diode (Ondax CP-405).
It has an in-built volume holographic grating to ensure wavelength stability over a large temperature range. 
Signal and idler photons are separated by a dichroic mirror and detected by silicon avalanche photodiodes (APD). 

The downconversion process is carried out in a {\it $\beta$}-barium-borate (BBO) crystal (5$\times$5$\times$5\,mm$^{3}$) cut for type I phase matching at $28.78^{\circ}$.
An advantage of using BBO in low resource systems is its relative tolerance to changes in temperature - this is useful aboard CubeSats where fine temperature control is a challenge.
The designed angle cut enables collinear pair production of horizontally polarized photon pairs at 760 (signal) and 867\,nm (idler).
Excess pump light is removed via appropriate filters.
Signal and idler photons are separated by a dichroic mirror (transmits idler) and detected by APDs (Laser Components SAP500). 
A mirror folds the optical path to satisfy the size constraint imposed by the nanosatellite platform.

Polarization correlation between signal and idler is measured using a combination of liquid crystal (LC) polarization rotators and polarizing beam splitter (PBS) cubes.
Using LC devices minimizes power and space requirements. 
It also reduces undesirable torque that may destabilize the orientation of the nanosatellite.
The polarization rotators are wavelength optimized for our source and have an extinction ratio larger than 100:1.
The PBS cubes provide redundancy in the detectors by allowing the use of two pairs of APDs.
In any single experiment, a fixed pair of APDs is used (1\&4 or 2\&3). 

A tool-kit for precise alignment of crystals within the optical unit was developed. 
Once the crystals are aligned they are secured to the optical unit by epoxy. 
The optical unit is made light-tight by opaque polyacetal (Delrin) walls containing all laser emission.
The complete optical unit (Fig.\,\ref{opticssetup}(b)) weighs less than 120\,gm.  The optical design has sufficient space for placement of additional crystals to convert this source into an entangled photon pair source \cite{trojek08}. 

The LC rotators are calibrated before insertion into the optical unit. 
When performing a single data run for APDs 1\&4, the LC device for the idler photon is supplied with a voltage that enables maximum transmission of horizontally polarized photons to the idler detector (APD1).
The rotator for the signal photon is then stepped through a series of voltages that correspond to a full polarization rotation of 2$\pi$. 
The rotator is allowed to stabilize at each new voltage step for 0.3\,s before data is collected.
When operating the APDs 2\&3 the same protocol is used except that the idler LC device is tuned to maximize transmission to APD3. 

\begin{figure}[ht]
	\begin{center}
	\includegraphics[scale=0.85]{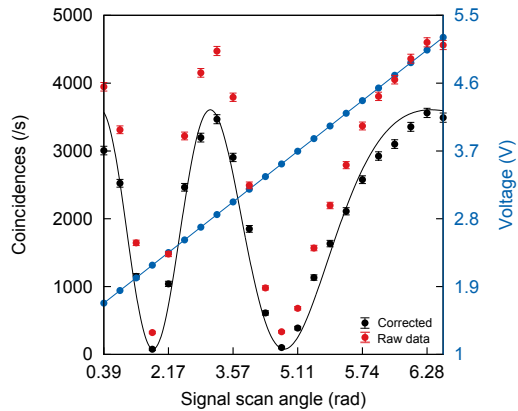}
	\caption{Typical data set for one complete measurement where the polarization of the signal photons is rotated over $2\pi$ by linearly stepping up the voltage to the liquid crystal rotator. A retarded sinusoid fitted to the corrected data has a visibility (contrast) of $95 \pm 1\%$. The retardation is due to the nonlinear response of the liquid crystal to voltage.   
	} 
	\label{singlevisibility}
	\end{center}
	\end{figure}

A typical cycle for APDs 1\&4 is shown in Fig.\,\ref{singlevisibility} when the pump is operated at 9\,mW.
Approximately 4500 photon pairs are detected when the signal and idler count rates are 360,000 and 330,000 respectively.
The maximum corrected coincidence event is about 3600 per second (coincidence time window is 9\,ns).

A sinusoid can be fitted to the corrected correlation data. 
The visibility (contrast) obtained from the sinusoid indicates the quality of the polarization correlations.
In Fig.\,\ref{singlevisibility}, the visibility from the corrected data is $95\pm 1\%$.
The time taken to complete the entire data collection is less than 30\,s.
This design enables photon pair sources to undergo fast and low power self-certification of correlation quality.

The existing design has a pair-to-singles ratio of approximately 1\%.
The largest contribution to this value is the poor overlap (39\%) of the detector active area (dia. 0.5\,mm) with the collimated pump mode (dia. 0.8\,mm). 
Detector inefficiency and imperfect transmission of devices (e.g. LC rotators) account for most of the remaining loss. 
Overlap between detector and pump mode can be greatly increased with the use of collection optics but is not implemented in the current design due to form factor limits. 
CubeSats sizes larger than 1U can relax the restriction but for increased launch opportunities we have designed for the smallest nanosatellite.

\section{The Electronic Sub-system}
\begin{figure}[htbp] \begin{center} \includegraphics[scale=0.9]{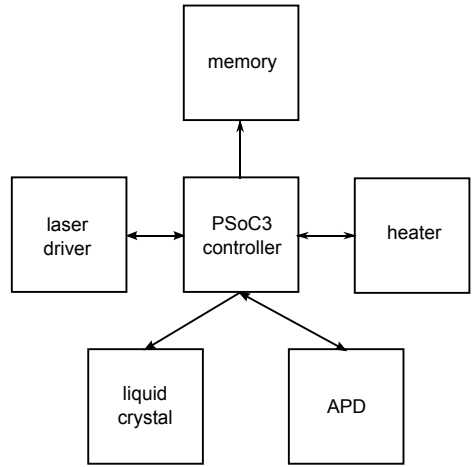} \caption{Modules of the control system are built around the micro-controller embedded in a Programmable System-on-Chip. The direction of arrows indicates communication flow. The laser driver, APD and heating modules are operated with feedback control. Data generated is stored on external flash memory for further processing.} \label{eleflow} \end{center} \end{figure}

The electronic sub-system is built around a Programmable System-on-Chip (Cypress PSoC3). 
The advantage of using a PSoC3 design is that it provides a micro-controller integrated with a large number of mixed signal processing functions.
This reduces design complexity and power consumption while maintaining functionality.
The embedded micro-controller enables rapid development of the automation software needed to operate the photon pair source and data collection.
The integrated electronics and optical system has a total mass below 250\,gm.  
A simplified block diagram illustrating the connections between the PSoC3 and various major modules is presented in Fig.\,\ref{eleflow}. 

The system is fully autonomous.
Upon power-up the PSoC3's first task is to monitor the temperature of the optical housing via thermistors which are part of the heater module. 
Operating temperature is set between 20$^{\circ}$C and 30$^{\circ}$C as the device is assembled in that range. 
When the temperature is outside this range, two options are available. 
If it is too warm, the PSoC3 will wait until the temperature falls into range. 
Otherwise the heater is activated to warm up the optical housing. 
Analysis of nanosatellite orbits indicate that CubeSat payloads in LEO tend to fluctuate approximately between $-5^{\circ}$C and 20$^{\circ}$C in a 100 minute cycle \cite{kataoka10} so heating is likely to be necessary.

Once the housing has reached the desired temperature, the PSoC3 activates the laser driver. 
Typically, the diode maintains single mode output and stable power between 5$^{\circ}$C and 30${^\circ}$C. However, very infrequently, it is observed that mode hopping and power fluctuations arise when temperature changes rapidly. In the event of large power instability, the PSoC3 halts data collection and may activate the heater.
Normal operation resumes after the laser power stabilizes. 

The liquid crystal rotators and APDs have dedicated control circuits. In particular, a novel method for operating the APD was developed \cite{chuan13}.
In this method, the detection efficiency of the APD is fixed by keeping the avalanche signal strength at a constant amplitude.
The avalanche amplitude in turn is adjusted via the bias voltage supplied to the APD (typically 100-130\,V).
This enables the APD to be operated without fine temperature control which can be power intensive.
There is also no need to rely on lab-based calibration that may not apply after the APDs have been exposed to radiation in LEO.
Using the novel circuit, each APD requires approximately 0.25\,W over a 40$^{\circ}$C temperature range.
 
All data are stored in two separated sectors within the 2\,MB flash memory (Numonyx M25P80) for redundancy. 
Given 8 minutes of uninterrupted operation, the device generates approximately 128\,kB of data. 
This volume of data can be downloaded from a CubeSat in LEO to a ground station in a single pass if the the radio link operates at 1.25\,kBps in the ultra-high frequency spectrum.

To meet the power budget of 2\,W, we have tuned our components for optimum efficiency. 
Power usage per module is listed in Table\,\ref{powerbudget}. Total continuous power usage without active heating is 1.3\,W. 

\begin{table}[ht] \centering \begin{tabular}{c | c} \noalign{\smallskip} \hline \hline Sub-systems & Power consumption (W) \\ \hline \hline 2 APDs & 0.5  \\ \hline laser driver & 0.45   \\ \hline PSoC3 + memory  & 0.3  \\ \hline liquid crystal  & 0.05  \\ \hline heater & 0.7 (1.7)  \\ \hline \hline \noalign{\smallskip} \end{tabular} \caption{Power consumption breakdown for all electronic components. Power required (w/o heating) is approximately 1.3\,W. When only the PSoC3 is active, the heater may receive up to 1.7\,W.} \label{powerbudget} \end{table} 

\section{Qualification Tests}
To demonstrate the ruggedness of the device its tolerance to vacuum, temperature and vibration was investigated.
A baseline for the polarization correlation was established and all tests were performed on a single copy of the device.
The visibility of the polarization correlations before and after each qualification test is presented in Table\,\ref{vacthermal}. 
The radiation tolerance of APDs \cite{chuan13} and other electronic components \cite{chuan14} were also studied.

The first test investigated the tolerance of the device under vacuum. 
A source of concern is that outgassing under low pressure (e.g. from epoxy) may result in deposition of material on critical surfaces resulting in reduced optical transmission and quality.
In the first set of qualification tests laser diodes were operated over 1 hour at a pressure of $1 \times 10^{-4}$ mbar to demonstrate compatibility with vacuum conditions.
Once this was achieved the complete system was subjected to a vaccuum pressure of $1 \times 10^{-7}$ mbar over 24 hours.
The system was then removed and operated to check for degradation in optical strength and quality.

The tolerance of the device to thermal fatigue was also investigated by subjecting it to a temperature profile varying between $-10^{\circ}$C to $40^\circ$C in a 100 minute cycle for over 24 hours. 
This is a larger temperature range than expected on nanosatellites in LEO \cite{kataoka10}. 

The device was then subjected to a random vibration profile simulating launch conditions recommended by the most common rockets used to launch CubeSats \cite{cyclone4, rama98}.
The frequency range covered 20- 2000\,Hz and subjected each axis to over 7.4\,g (rms) of acceleration for at least 3 minutes.
The device was also subjected to a sine sweep in all axes to ensure that no mechanical resonance existed in the 5-100\,Hz range where typical rockets can impart large forces at these frequencies to the nanosatellites.

The polarization correlation visibility after each test is tabulated in Table II.
The stable performance of the correlated photon pair system under different test profiles suggest that it is possible to build a robust and small entangled photon source using similar techniques.

\begin{table}[ht]
\centering
\begin{tabular}{c|c|c|c}
\hline\hline
Test & vacuum & thermal & vibration (3-axis avg.) \\ \hline \hline
Visibility & $94 \pm 1$ & $93\pm 2$ & $92\pm 2$ \\ \hline \hline 
\end{tabular}
\caption{Polarization correlation visibility after qualification tests. The baseline visibility before the tests was $93\pm2$\%.} \label{vacthermal} 
\end{table}

\section{Near-space Demonstration}

To investigate the performance of the system close to the target operating environment (400\,km LEO) a helium filled weather balloon was used to lift the device into near-space.
In addition to the device, the balloon carried tracking and telemetry instrumentation such as a GPS receiver that broadcast its position on an amateur radio band (144.8\,MHz) at all times.
Sensors recording 3-axis acceleration, air temperature, relative humidity and pressure were packaged with the correlated system, controlled by a separate micro-controller.
The overall mass of the balloon package was around 2\,kg.

\begin{figure}[ht] \begin{center} \includegraphics[scale=0.4]{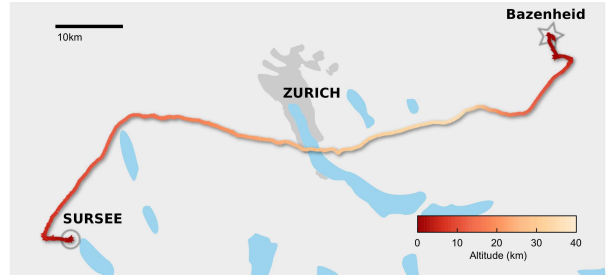} \caption{Flight path of the helium balloon hosting the device under test. The balloon reached a maximum altitude of 35.5\,km before bursting. } \label{map} \end{center} \end{figure}

\begin{figure}[h!] \begin{center} \includegraphics[scale=0.7]{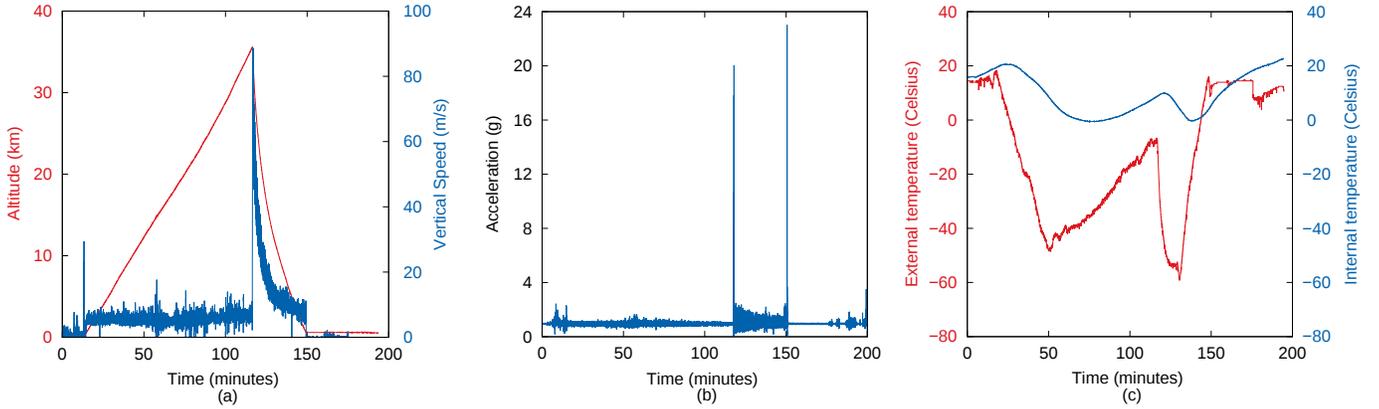} \caption{Conditions experienced by the photon pair system during the test flight. The system was tested up to 35.5\,km (a) which is 10\% of the planned space mission altitude. Acceleration events over 20 g (b) were experienced when the helium balloon burst and the package landed. Increased acceleration due to tumbling is observed during the descent. The troughs of the temperature profiles (c) correspond to the balloon ascending and descending through the jet stream where it reached horizontal speeds up to 32\,ms$^{-1}$.} \label{grandballoon} \end{center} \end{figure}

The device was activated on the ground 15 minutes before the balloon was released.
The release point was in Sursee, Switzerland (N\,47.17$^\circ$, E\,8.1$^\circ$) and landed at Bazenheid, Canton St. Gallen 130\,km away (N\,47.42$^\circ$, E\,9.1$^\circ$) 
The flight path is plotted in Fig.\,\ref{map} and environmental data from the flight are plotted in Fig.\,\ref{grandballoon}.
The test flight took around 2 hours and reached a ceiling attitude of 35.5\,km above sea level before bursting. 
The rate of ascent was approximately $5\,\text{ms}^{-1}$ and after the balloon burst the maximum descent velocity was over $90\,\text{ms}^{-1}$.
When the air density is sufficiently high, an attached parachute slowed the descent to below $10\,\text{ms}^{-1}$.

Total system acceleration experienced by the device is plotted in Fig.\,\ref{grandballoon}(b). 
The accelerometer (Analog Devices ADXL345) measures up to 16\,g per axis with 13-bit resolution.
The early acceleration events are due to the initial drag from the weather balloon during take off. 
Two major acceleration events occur after 120 and 150 minutes corresponding to the balloon burst (20\,g) and landing (23\,g) respectively, in an otherwise uniform plot. 

The internal and external temperature profiles are correlated (Fig.\,\ref{grandballoon}(c)).
The troughs of the temperature profiles corresponds to the balloon ascending and descending through the jet stream. 
The internal temperature varies from 0$^{\circ}$C and 15$^{\circ}$C in a low pressure environment (below 1\,mbar at 35.5\,km altitude).

\begin{figure}[ht] \begin{center} \includegraphics[scale=0.65]{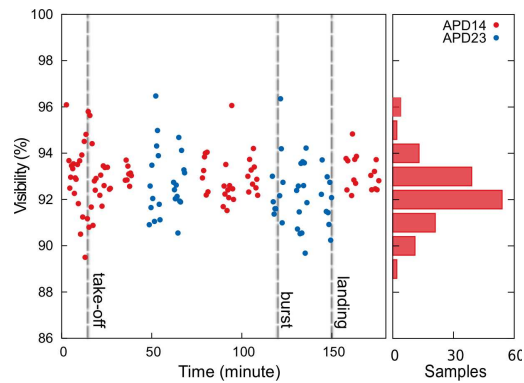} \caption{Polarization correlations during test flight. The visibility of the correlations is 92\% (within the error of laboratory demonstrations).} \label{sum} \end{center} \end{figure}

During flight the APD pairs were operated alternately so that the performance for all detectors could be monitored.	
The average visibility throughout the experiment is around 92\% (Fig\,\ref{sum}).
This near-space test flight demonstrates that the system continues to generate and monitor polarization correlations despite significant variation in acceleration, pressure and temperature. 

\section{Conclusion}

A compact, robust and power-efficient system for generating and monitoring polarization correlated photon pairs has been designed and assembled.
A single device was subjected to qualification tests that examined the system tolerance to vaccuum, temperature and vibration.
It was then placed on a weather balloon and lifted 35.5\,km into the upper atmosphere to test its performance under near-space conditions.
 
The test results demonstrate that the integrated source and detector package survives conditions that most laboratory-based setups cannot tolerate. 
The ultimate aim is to operate the device at an LEO altitude of about 400\,km. 
The balloon test has enabled the device to be tested to almost 10\% of that altitude. 
The combined suite of tests demonstrate that the system is at a high level of technological readiness paving the way for space-based demonstration of entangled photon sources.

In future work additional crystals will be integrated so that the source generates polarization entangled photon pairs. Different SPDC materials may be integrated into the device with minimal changes to the form factor and power requirement. 

In overcoming the resource limitations encountered on nanosatellites our design enables photon pair sources to be deployed over a larger number of low-resource operating environments.
It is anticipated that they will be deployed not only on satellites, but also on small unmanned aerial vehicles or even for handheld communicators.
The integration of measurement apparatus for monitoring the polarization correlations also serve as a self-diagnostic tool for analyzing the status of the source.
Rugged and efficient photon pair sources capable of self-diagnosis are highly desirable as nodes in {\it any} quantum network whether in space or on Earth.

\begin{acknowledgements}
We thank Tan Yue Chuan for his assistance in preparing the test devices and James Grieve for manuscript assistance. Invaluable assistance was provided by Jens Spinner and the Radio Club Sursee in tracking and recovering the balloon test package. The CQT team also thanks Daniel Oi for technical discussions regarding CubeSat technology. C. Cheng is supported by a DSO-CQT(NUS) grant.
\end{acknowledgements}

\bibliographystyle{naturemag}
\bibliography{tang14}

\end{document}